\documentstyle[12pt]{article}
\def\Re{{\cal R \mskip-4mu \lower.1ex \hbox{\it e}\,}}
\def\Im{{\cal I \mskip-5mu \lower.1ex \hbox{\it m}\,}}
\def\ie{{\it i.e.}}
\def\eg{{\it e.g.}}

\def\etal{{\it et al.}}

\def\sub#1{_{\lower.25ex\hbox{$\scriptstyle#1$}}}
\def\sul#1{_{\kern-.1em#1}}
\def\sll#1{_{\kern-.2em#1}}
\def\sbl#1{_{\kern-.1em\lower.25ex\hbox{$\scriptstyle#1$}}}
\def\ssb#1{_{\lower.25ex\hbox{$\scriptscriptstyle#1$}}}
\def\sbb#1{_{\lower.4ex\hbox{$\scriptstyle#1$}}}

\def\slash{\not\!\!}

\def\to{\rightarrow}
\def\mh{\ifmmode m\sbl H \else $m\sbl H$\fi}
\def\mch{\ifmmode m_{H^\pm} \else $m_{H^\pm}$\fi}
\def\mt{\ifmmode m_t\else $m_t$\fi}
\def\mc{\ifmmode m_c\else $m_c$\fi}
\def\mz{\ifmmode M_Z\else $M_Z$\fi}
\def\mw{\ifmmode M_W\else $M_W$\fi}
\def\mws{\ifmmode M_W^2 \else $M_W^2$\fi}
\def\mhs{\ifmmode m_H^2 \else $m_H^2$\fi}
\def\mzs{\ifmmode M_Z^2 \else $M_Z^2$\fi}
\def\mts{\ifmmode m_t^2 \else $m_t^2$\fi}
\def\mcs{\ifmmode m_c^2 \else $m_c^2$\fi}
\def\mchs{\ifmmode m_{H^\pm}^2 \else $m_{H^\pm}^2$\fi}
\def\ztwo{\ifmmode Z_2\else $Z_2$\fi}
\def\zone{\ifmmode Z_1\else $Z_1$\fi}
\def\mtwo{\ifmmode M_2\else $M_2$\fi}
\def\mone{\ifmmode M_1\else $M_1$\fi}
\def\tb{\ifmmode \tan\beta \else $\tan\beta$\fi}
\def\xw{\ifmmode x\sub w\else $x\sub w$\fi}
\def\ch{\ifmmode H^\pm \else $H^\pm$\fi}
\def\lum{\ifmmode {\cal L}\else ${\cal L}$\fi}
\def\inpb{\ifmmode {\rm pb}^{-1}\else ${\rm pb}^{-1}$\fi}
\def\infb{\ifmmode {\rm fb}^{-1}\else ${\rm fb}^{-1}$\fi}
\def\epem{\ifmmode e^+e^-\else $e^+e^-$\fi}
\def\ppb{\ifmmode \bar pp\else $\bar pp$\fi}

\newskip\zatskip \zatskip=0pt plus0pt minus0pt
\def\matth{\mathsurround=0pt}

\def\atversim#1#2{\lower0.7ex\vbox{\baselineskip\zatskip\lineskip\zatskip
  \lineskiplimit 0pt\ialign{$\matth#1\hfil##\hfil$\crcr#2\crcr\sim\crcr}}}

\renewcommand{\thefootnote}{\fnsymbol{footnote}}

\begin{document} \begin{titlepage}
\setcounter{page}{1}
\thispagestyle{empty}
\rightline{\vbox{\halign{&#\hfil\cr
&ANL-HEP-PR-93-87\cr
&November 1993\cr}}}
\vspace{1in}
\begin{center}

{\Large\bf
Model Dependence of $W_R$ Searches at the Tevatron}
\footnote{Research supported by the
U.S. Department of
Energy, Division of High Energy Physics, Contracts W-31-109-ENG-38.}
\medskip

\normalsize THOMAS G. RIZZO
\\ \smallskip
High Energy Physics Division\\Argonne National
Laboratory\\Argonne, IL 60439\\

\end{center}

\begin{abstract}

We explore the sensitivity of the on-going Tevatron search for charged,
right-handed gauge bosons, $W_R^{\pm}$, to various model dependent
assumptions such as the magnitude of the $SU(2)_R$ gauge coupling, the
values of the right-handed Kobayashi-Maskawa mixing matrix elements,
$(V_R)_{ij}$, and the nature of the right-handed neutrino. These results
also have important implications for HERA searches for right-handed currents.

\end{abstract}

%\vskip1.75in

%\noindent{(Talk given at the {\it Workshop on Photon Radiation from Quarks},
%Annecy, France, December 2-3, 1991.)}

\renewcommand{\thefootnote}{\arabic{footnote}} \end{titlepage}

%%%%%%%%%%%%%%%%%%%%%%%%%%%%%%%---- text

Despite the many successes of the Standard Model(SM), there are many reasons
to believe that new physics must exist at a scale not far above that being
probed by current accelerator experiments. These beliefs originate from the
fact that too many of the pieces of the SM are put in by hand in order to
conform to experimental observation. Perhaps one of the oldest of these
pieces is the $V-A$ nature of the charged current interaction which forces
the SM gauge group to be its canonical $SU(2)_L \times U(1)_Y$ structure.
One of the earliest extensions of the SM, the Left-Right Symmetric Model(LRM)
{\cite {lrm}}, which is based on the gauge group
$SU(2)_L \times SU(2)_R \times U(1)$,
`explains' the apparent absence of right-handed currents(RHC)
by associating them with a much more massive gauge boson, $W_R^{\pm}$. This
model, in its more modern, supersymmetric version can be nicely embedded into
an $SO(10)$ GUT structure which yields correct predictions for
$sin^2 \theta(M_Z)$ and $\alpha_s(M_Z)$, interesting relationships among
neutrino masses, and allows for the possibility that $W_R$ can be lighter than
a few TeV{\cite {newlrm}} and hence potentially visible at existing or planned
colliders. In this paper we would like to focus upon several specific aspects
of this model related to the direct
searches for $W_R$ at the Tevatron. As we will see, these considerations will
have important implications for the RHC searches at HERA as well.

In order to establish limits on the mass of $W_R$'s, either from low energy
data or from collider searches, there are five important aspects of the LRM
which come into play which can be phrased as a series of questions:

($i$) How large is the ratio of the $SU(2)_R$ and $SU(2)_L$ coupling constants,
$\kappa=g_R/g_L$? Ordinarily one might expect such a ratio to be of order
unity and it can be shown{\cite {us}} that internal consistency within
the LRM requires
that $\kappa^2 \geq x_w/(1-x_w)$, where $x_w=sin^2 \theta_w$. Numerically,
this implies $\kappa \geq 0.55$. With the GUT context, however, $\kappa$ is
either very close to unity or lies in the range $0.55 \leq \kappa \leq 1$.
The implication of the size of $\kappa$ for $W_R$ Tevatron collider searches is
quite obvious as the production cross section is quadratic in $\kappa$. Thus
as $\kappa$ decreases(increases) in magnitude the resulting $W_R$
search reach is reduced(enhanced).

($ii$) What is the magnitude of the mass of the right-handed neutrino, $\nu_R$?
Clearly, if neutrinos are Dirac fields then $\nu_R$ is simply a part of the
four component $\nu$ spinor and is thus essentially massless. However, if the
rather attractive see-saw mechanism{\cite {seesaw}} is invoked, $\nu_R$ is a
heavy Majorana neutrino. If the Dirac path is realized, the lightness of the
neutrinos imply that they appear as missing $E$ or $p_t$ in collider detectors
and that polarized $\mu$ decay experiments{\cite {mu}} can place stringent
limits on the $W_R$ mass, of order 480 GeV, as well as its possible
mixing with the SM $W$. If, however,
$\nu_R$'s are heavy this situation changes drastically. For example, if
$\nu_R$'s are more massive than a few hundred MeV then they cannot be
produced as final states in $K$, $\pi$, or $\mu$ decay thus avoiding the low
energy bounds. If $\nu_R$'s are even heavier, they can easily decay inside the
collider detector and the missing $E$ or $p_t$ signature is lost. The
resulting final state would then consist of two leptons plus two jets with
only one of the leptons being isolated and at very high $p_t$. Depending
on the $\nu_R$ mass,
the second lepton and both jets may be quite close in $\Delta R$. Such a
scenario would require a completely different search technique than what is
conventionally employed and is outside of the scope of the present paper.

($iii$) What is the branching fraction($B$) for leptonic $W_R$ decays? Since
conventional Tevatron searches require the presence of a hi-$p_t$ lepton, a
reduction in the value of $B$ due to the existence of $W_R$ decays into non-SM
final states, such as SUSY particles, will result in a loss of mass reach.

($iv$) Perhaps the most important and least easily addressed question is `what
are the values of the elements of the `right-handed'
Kobayashi-Maskawa(KM) mixing
matrix, $V_R$?' Most analyses of the LRM assume that the elements of $V_R$ and
the conventional KM matrix, $V_L$, differ at most by phase factors. If
$V_R=V_L$, then it has been known for some time that considerations of
the $K_L-K_S$ mass
difference result in a strong lower bound{\cite {klks}} on the mass of $W_R$ of
1.6 TeV thus placing it outside the search capabilities of existing
colliders. If,
however, we remove the constraint of $V_R=V_L$ and allow $V_R$ to be arbitrary,
even in the absence of fine-tuning we find that $W_R$ can be as light as 280
GeV for a top quark mass of 160 GeV. Also if $V_R$ differs from $V_L$
significantly, the $W_R$ production cross section at the Tevatron
can be drastically reduced
since the initial valence $u\bar d$ parton flux has the greatest luminosity.
It is important to note that HERA searches for RHC are not susceptible to this
$V_R$ uncertainty. Consider the scattering of $e^-_R$ off of valence
u-quarks via
$W_R^-$ exchange. At the parton level, depending on the form of $V_R$, the
initial u-quark is transformed mostly into d-, s-, or b-quarks. However, if we
sum over all three final states and neglect the b-quark mass as a first
approximation, we find the
resulting cross section to be {\it independent} of $V_R$ due to the fact that
$V_R$ is unitary. This implies that the usually quoted search reach for $W_R$
at HERA{\cite {hera}}, using right-handed polarized $e^-$ beams, of
approximately 400 GeV (assuming $\kappa=1$ and light $\nu_R$'s) is quite
insensitive to the form of $V_R$. We note, however, that the corresponding
result for $e^+_R$ may be reasonably $V_R$ sensitive since in this
case the initial
valence d-quark can be transformed into u-, c-, or t-quarks. Since top quarks
are quite massive, their production is highly suppressed so that we can no
longer make use of the unitarity argument above and the possibility of strong
$V_R$ dependence in this channel remains. We note that if $\nu_R$'s are
sufficiently massive as to decay inside a HERA detector, the game is totally
different as the SM background is now drastically reduced. It has in fact
been shown by Buchm\"uller \etal{\cite {hera}} that the $W_R$ search range is
significantly enhanced (to over 700 GeV for 120 GeV $\nu_R$'s) in this case.

($v$) A last question one might ask is `what is the mass of the $Z'$
associated with the $W_R$ in the LRM?' In general, the masses of these two
particles are related, in the absence of mixing, via the expression{\cite {us}}
\medskip
\begin {equation}
{M_{W_R}^2\over {M_{Z'}^2}} =  {{(1-x_w)\kappa^2-x_w}\over {\rho_R(1-x_w)
\kappa^2}}
\end {equation}
\medskip
where the parameter $\rho_R$ takes on the value 1(2) if the $SU(2)_R$
breaking sector consists solely of Higgs doublets(triplets). (The triplet
scheme is favored in the see-saw scenario for neutrino masses.) From this we
see that unless the $SU(2)_R$ breaking sector is somewhat unusual, the
$Z'$ will always be more massive that the $W_R$. While $W_R$ search limits
may be sensitive to $V_R$, however, those for $Z'$ are not although they too
are subject to uncertainties in $\kappa$ and the $Z'$ leptonic branching
fraction. For $\kappa=1$ and $Z'$ decays to SM fermions only, the CDF published
limit{\cite {cdf}}
from the 1988-89 run of 412 GeV on a $Z'$ with SM couplings would translate
into a indirect, but
$V_R$-independent, lower limit on $M_{W_R}$ of only 302(214) GeV for
$\rho_R=1(2)$. An incomplete analysis of the CDF electron data from run Ia
places the corresponding lower limit of 495 GeV on a $Z'$ with SM couplings
would imply the $V_R$-independent lower limit on $M_{W_R}$ of 371(263) GeV
for $\rho_R=1(2)$.
While these results are instructive the bounds we obtain are relatively
weak and could be significantly loosened if $\rho_R$ were greater than 2 and/or
the $Z'$ leptonic branching fraction was suppressed.

The strongest published bound on the $W_R$ mass from direct Tevatron
searches is that
of the CDF Collaboration{\cite {cdf} obtained from their 1988-89 data by
combining their electron and $\mu$ samples: $M_{W_R} \geq 520$ GeV. Their
analysis assumes HMRSB parton distributions{\cite {hmrs}}, $\kappa=1$,
$V_L=V_R$, $B=1/12$, with $M_{\nu_R}<15$ GeV and
$\nu_R$ appearing as $\slash E$ or $\slash p_t$. (The D0 Collaboration has
recently reported a corresponding preliminary limit of $M_{W_R} \geq 600$ GeV
from the 1992-93 Tevatron run Ia with essentially identical
assumptions{\cite {d0}}.) With
data from Tevatron run Ia currently being analyzed and the 1993-94 run Ib soon
to begin in earnest, we would like to address the issue of how these existing
limits, as well as the limits obtainable from the new data would be
modified if these assumptions are loosened. In what follows, we will
still assume that the $\nu_R$ is sufficiently light so that neither
the leptonic branching fraction nor the $\slash p_t$ signature are
significantly effected. (Of course, we still can take these $\nu_R$'s to be
sufficiently massive in order to avoid $\mu$-decay constraints
while maintaining their `stability' as
far as collider searches are concerned.)  For simplicity we will assume
$B$ to be directly obtainable
from a calculation including only SM final states once finite top-quark mass
and three-loop QCD corrections are applied. (To be definitive, we assume
$m_t=160$ GeV and take $\alpha_s(M_Z)=0.123$ which we then run up to $M_{W_R}$
using the
three-loop renormalization group equations.) We thus will address the
sensitivity of the Tevatron searches to variations in $\kappa$ as well as
$V_R$. In our analysis, all production cross sections will be calculated
assuming the CTEQ1M{\cite {cteq}} parton distribution functions as well as
a `K-factor' arising from QCD corrections{\cite {kfact}}.

Let us first deal with varying $V_R$ assuming $\kappa=1$; we will return to
the more sophisticated case below. In general, the elements of $V_R$ are
determined
by three angles and a number of phases. In order to demonstrate the sensitivity
of the Tevatron $W_R$ search to variations in $V_R$, it is sufficient to
assume only a single phase is present. We first generate a single set of these
parameters and calculate the absolute squares of the nine elements in $V_R$,
$|(V_R)_{ij}|^2$. We next calculate the parton level processes $q_i\bar q_j
\to W_R \to \ell \nu_R$ and weight them by the corresponding parton
luminosities evaluated at $Q^2=M_{W_R}^2$. When these are scaled by the
squares of the elements of $V_R$ and summed over $i,j$ a final total cross
section is obtained for a fixed $W_R$ mass. $M_{W_R}$ is then increased until
the experimental limiting value is reached. For the 1988-89 run, we use
the CDF limit curve
as presented in their paper{\cite {cdf}}. For runs Ia and Ib, we will simply
rescale this CDF curve by the corresponding ratios of the integrated
luminosities. This approximation does not allow, however, for improvements
in the detector acceptance or backgrounds analyses. Since we are more
interested in how the $W_R$ search reach {\it changes} as $V_R$ is varied we
feel this is a reasonable simplification for this kind of analysis.

The above
procedure needs to be repeated many times via a Monte Carlo so that an
adequate coverage of the $V_R$ parameter space volume is obtained. This can
be judged by increasing the number of generated points in this space by an
order of magnitude and observing the sensitivity of the resulting limits to
this variation. For a fixed value of $\kappa$, we find that $10^6$ points
proves to be quite adequate to cover the entire $V_R$
parameter space volume. Once the $W_R$ mass limit for each of the generated
points in the $V_R$ parameter space is
determined we cluster them in bins of 2 GeV and present the results as a
histogram over the $W_R$ mass. In an alternate approach, one can imagine
instead using the Monte Carlo to generate the squares of four of the elements
of $V_R$ and then using unitarity to obtain the others. This analysis would
then assume that the squares of the $V_R$ elements would have flat
distributions instead of the corresponding flat distributions for the angles
and phases themselves. The results of these two approaches would yield
qualitatively similar results.

Fig. 1a shows the results of this procedure for the CDF 1988-89 Tevatron
data sample
with $\kappa=1$. Several features of this figure, in addition to the rather
long tail to the left of the peak, are important to observe:
($i$) A reasonably large fraction of the `events' lie close to the upper end
of the distribution; in fact, 23.5$\%$ lie at or above 500 GeV. This means
that for a sizeable fraction of the parameter space volume the actual $W_R$
mass reach is not too much different than what would be obtained if $V_R=V_L$.
($ii$) 29.8$\%$ of the `events' lie below 400 GeV, the nominal HERA search
limit. This would imply, based only upon this set of data, that HERA may still
have a sizeable chance to be able to observe RHC even if $\nu_R$ is light.
($iii$) A statistically
significant enhancement is observed in the region near $M_{W_R}=360$ GeV. This
arises from a situation where $(V_R)_{us}$ is big and takes advantage of
the fact that
the $u\bar s$ parton luminosity is the second largest. ($iv$) Although it is
unlikely, there is a small chance, 0.61$\%$, that $M_{W_R}$ may lie at or
below 300 GeV.

The generic shape of this distribution persists for increased integrated
luminosities (as well as for different values of $\kappa$). Fig. 1b(c) shows
the corresponding results for run Ia(Ib); in the Ib case, an integrated
luminosity of 75 $pb^{-1}$ has been assumed. From Fig. 1b we see that if no
$W_R$ candidates are observed after the data is analyzed, the probability that
HERA can observe RHC (for the case of light $\nu_R$!) is still non-zero, but
quite small, \ie, only 0.23$\%$. Note that the distribution has elongated as
well as flattened and the `$u\bar s$' enhancement still persists near 470 GeV
although it appears to be somewhat smaller.
Increasing the luminosity further to the run Ib case(Fig.~1c) we see that
these general trends continue. At the 75 $pb^{-1}$ level, we see that there are
no events below about 460 GeV implying that RHC would not be observable at
HERA if the Tevatron data shows no hint of $W_R$ with this integrated
luminosity.

What happens when $\kappa \neq 1$? The case where $V_R=V_L$ is rather simple
and is shown in Fig.~2 where the mass reach for the three Tevatron runs is
plotted as a function of $\kappa$. Note that for $0.55 \leq \kappa \leq 1$,
which is the theoretically expected range, the mass reach can vary by as much
as 100 GeV. One possible way of dealing with arbitrary $\kappa$ is to present
results similar to the above for some representative values, \eg, in
Figs.~3a-c, we show what happens for $\kappa=0.85$. Essentially, to a first
approximation, all of the curves in Fig.~1 are simply shifted to the left,
\ie, to lower values of $W_R$.
As a second approach, taking the theoretical bias into account, we may imagine
treating $\kappa$ in the above range as a free parameter and placing it on an
equal footing with the various angles and phases in $V_R$ as part of the Monte
Carlo. To do this, we increase the number of points in the $V_R$ parameter
space by 2 and generate an equal number of $\kappa$ values for which we
also assume
a flat distribution. The result of this approach is shown in Figs.~4a-c for
the three Tevatron runs. Allowing $\kappa$ to vary within the parameter Monte
Carlo totally changes the shape of the anticipated $W_R$ mass reach
distribution resulting from `$\kappa$ smearing'. In addition to
the tail which goes down to rather low $M_{W_R}$ values, these figures
show two sizeable enhancements. The one at larger $M_{W_R}$
results from the case where $(V_R)_{ud}$ is large but $\kappa <1$ reduces the
limit from its maximum allowed value. In the case of the 1988-89 run, \eg,
the maximum search reach for large $(V_R)_{ud}$ is reduced, on average, about
60-70 GeV which explains the position of the peak. Note that the approximate
position of the peak relative to the largest $M_{W_R}$ value stays roughly
constant as the integrated luminosity is increased. The somewhat smaller peak
at lower $M_{W_R}$ is the result of the large $(V_R)_{us}$ possibility as well
as feed-down from the case of large $(V_R)_{ud}$ when $\kappa$ is close
to 0.55. We note that as the
integrated luminosity increases these two peaks separate and the one at larger
$M_{W_R}$ becomes more pronounced, although its height is not increased, while
the smaller one is reduced to being nearly a shoulder on the tail of the low
mass end of the distribution. This results in an increased skewness of the
mass reach distribution. Also, as the luminosity increases the apparent
width of these distributions change; we can see this by calculating the
average value and standard deviation of the $W_R$ mass reach for these
three cases. We find
$M_{W_R}=397.7\pm 62.2, 540.5\pm 76.4, 629.8\pm 82.1$ GeV for the 1988-89,
Ia, and Ib runs respectively.

As an application of the above analysis, we briefly consider the model of
Gronau and Wakaizumi(GW) in which b-quark decays occur only through the
exchange of $W_R$'s{\cite {gron} and $\nu_R$ is relatively light. Assuming
the form of $V_R$ as originally
suggested in their model, we can now calculate the Tevatron mass reach as a
function of $\kappa$ as shown in Fig.~5. The rather large values obtained
here can be easily traced back to the large size of $(V_R)_{ud}$ in this
scenario. Similarly, we can determine a
{\it {upper}} bound on the $W_R$ mass in their model by demanding agreement
with the most recent determination of $V_{cb}${\cite {last}}, which is also
shown in Fig.~5. Combining these two constraints we see that the 1988-89 CDF
Tevatron data forces $M_{W_R}>560$ GeV and $\kappa>1.35$ while the anticipated
results from run Ia will increase these limits to $M_{W_R}>750$ GeV and
$\kappa>1.85$ assuming no signal events are observed. One may argue that
although such large values of $\kappa$ may be {\it a priori} allowed, they
are perhaps unnaturally large and are certainly outside of the range
anticipated
in grand unified models. Clearly, data from Tevatron run Ib would only
push both these
quantities to even higher values assuming no signal events are observed. We
may conclude from these considerations that for this model to remain viable
a different form of $V_R$ must be
assumed than what was originally suggested.

In summary, we have analyzed the sensitivity of Tevatron searches for $W_R$ to
various assumptions about the parameters of the LRM, in particular, the
value of $\kappa$ and the form or the right-handed mixing matrix, $V_R$.
Hopefully, $W_R$ will be sufficiently light as to been observed in the next
round of collider experiments.

\vskip.25in
\centerline{ACKNOWLEDGEMENTS}

The author would like to thank J.L.\ Hewett, M.\ Gronau, H.\ Contopanagos,
S.\ Moulding(CDF Collaboration), N.\ Hadley(D0 Collaboration), and the
various members of the LSGNA Collaboration for fruitful discussions. This
research was supported in part by the U.S.~Department of Energy under
contract W-31-109-ENG-38.

\newpage

%
%%%%%%%%%%%%%%%%%%--- References
%%%%%%%%%%%%%%%%%%%%%%%%%%%%%%%%%%%%%%%%%%%%%%%%%%%%%%%
\def\MPL #1 #2 #3 {Mod.~Phys.~Lett.~{\bf#1},\ #2 (#3)}
\def\NPB #1 #2 #3 {Nucl.~Phys.~{\bf#1},\ #2 (#3)}
\def\PLB #1 #2 #3 {Phys.~Lett.~{\bf#1},\ #2 (#3)}
\def\PR #1 #2 #3 {Phys.~Rep.~{\bf#1},\ #2 (#3)}
\def\PRD #1 #2 #3 {Phys.~Rev.~{\bf#1},\ #2 (#3)}
\def\PRL #1 #2 #3 {Phys.~Rev.~Lett.~{\bf#1},\ #2 (#3)}
\def\RMP #1 #2 #3 {Rev.~Mod.~Phys.~{\bf#1},\ #2 (#3)}
\def\ZP #1 #2 #3 {Z.~Phys.~{\bf#1},\ #2 (#3)}
\def\IJMP #1 #2 #3 {Int.~J.~Mod.~Phys.~{\bf#1},\ #2 (#3)}

\newpage

%%%%%%%%%%%%%%%%%%%%%%%--- figures
%
{\bf Figure Captions}
\begin{itemize}

\item[Figure 1.]{Histogram of the $W_R$ mass reach at the Tevatron assuming
$\kappa=1$ employing the CTEQ1M parton distributions as well as a `K-factor'
from QCD corrections. Results are shown for the 1988-89 Tevatron run(a), as
well as for Tevatron runs (b) Ia and (c) Ib. In the run Ib case, an integrated
luminosity of 75 $pb^{-1}$ is assumed.}
\item[Figure 2.]{Mass reach as a function of $\kappa$ for the 1988-89 Tevatron
run(dots) as well as for run Ia(dash) and run Ib(dash-dots) assuming that
$V_R=V_L$.}
\item[Figure 3.]{Same as Fig.~1 but for $\kappa$=0.85.}
\item[Figure 4.]{Same as Fig.~1 but now $\kappa$ is allowed to vary within
the Monte Carlo along with the elements of $V_R$ over the
range $0.55 \leq \kappa \leq 1$ in accordance with theoretical expectations.}
\item[Figure 5.]{Mass reach as a function of $\kappa$ for the 1988-89 Tevatron
run(dots) as well as for run Ia(dash) assuming that $V_R$ takes the form as
given by the Gronau and Wakaizumi model. The solid line is the 95$\%$ CL upper
bound on the $W_R$ mass in their model.}
\end{itemize}

\end{document}